\documentclass[conference,10pt]{IEEEtran}
\IEEEoverridecommandlockouts

\usepackage{cite}
\usepackage{amsmath,amssymb,amsfonts}
\usepackage{graphicx}
\usepackage{textcomp}
\usepackage{xcolor}
\def\BibTeX{{\rm B\kern-.05em{\sc i\kern-.025em b}\kern-.08em
		T\kern-.1667em\lower.7ex\hbox{E}\kern-.125emX}}

\usepackage{subfigure}
\usepackage{epstopdf}

\newtheorem{lemma}{Lemma}
\newtheorem{proposition}{Proposition}
\newtheorem{remark}{Remark}

\makeatletter
\def\endthebibliography{%
	\def\@noitemerr{\@latex@warning{Empty `thebibliography' environment}}%
	\endlist
}
\makeatother

\begin{document}

\include{header}

\title{Transmit Optimization for Multi-functional MIMO Systems Integrating Sensing, Communication, and Powering}

\author{\IEEEauthorblockN{Yilong Chen, Haocheng Hua, and Jie Xu}
	\IEEEauthorblockA{School of Science and Engineering and Future Network of Intelligence Institute, \\
	The Chinese University of Hong Kong (Shenzhen), Shenzhen, China \\
	Email: yilongchen@link.cuhk.edu.cn, haochenghua@link.cuhk.edu.cn, xujie@cuhk.edu.cn}
	\thanks{J. Xu is the corresponding author.}
}

\maketitle

\begin{abstract}
	This paper unifies integrated sensing and communication (ISAC) and simultaneous wireless information and power transfer (SWIPT), by investigating a new multi-functional multiple-input multiple-output (MIMO) system integrating wireless sensing, communication, and powering. In this system, one multi-antenna hybrid access point (H-AP) transmits wireless signals to communicate with one multi-antenna information decoding (ID) receiver, wirelessly charge one multi-antenna energy harvesting (EH) receiver, and perform radar sensing for a point target based on the echo signal at the same time. Under this setup, we aim to reveal the fundamental performance tradeoff limits of sensing, communication, and powering, in terms of the estimation Cram{\'e}r-Rao bound (CRB), achievable communication rate, and harvested energy level, respectively. Towards this end, we define the achievable CRB-rate-energy (C-R-E) region and characterize its Pareto boundary by maximizing the achievable rate at the ID receiver, subject to the estimation CRB requirement for target sensing, the harvested energy requirement at the EH receiver, and the maximum transmit power constraint at the H-AP. We obtain the semi-closed-form optimal transmit covariance solution to the formulated problem by applying advanced convex optimization techniques. Numerical results show the optimal C-R-E region boundary achieved by our proposed design, as compared to the benchmark scheme based on time switching.
\end{abstract}


\section{Introduction}

Future sixth-generation (6G) wireless networks need to incorporate billions of low-power Internet-of-things (IoT) devices and support their localization, sensing, communication, and control in a sustainable manner. Towards this end, simultaneous wireless information and power transfer (SWIPT) \cite{clerckx2019fundamentals} and integrated sensing and communications (ISAC) \cite{liu2022integrated} have emerged as two enabling techniques aiming to reuse radio signals to wirelessly charge massive IoT devices and provide sensing functionality, respectively. With their recent advancements, we envision that future 6G networks will integrate both SWIPT and ISAC to evolve towards new multi-functional wireless systems, which can provide sensing, communication, and powering capabilities at the same time. Such multi-functional wireless systems are expected to significantly enhance the utilization efficiency of scarce spectrum resources and densely deployed base station (BS) infrastructures, and facilitate the localization and powering of massive low-power devices to support emerging IoT applications.

In the literature, there have been extensive prior works investigating the transmit optimization for SWIPT (e.g., \cite{zhang2013mimo, xu2014multiuser}) and ISAC (e.g., \cite{xiong2022flowing, hua2022mimo, liu2020joint, hua2021optimal}) independently. For instance, the authors in \cite{zhang2013mimo} first studied a multiple-input multiple-output (MIMO) SWIPT system with one information decoding (ID) receiver and one energy harvesting (EH) receiver, in which the transmit covariance at the BS was designed to optimally balance the tradeoff between the communication rate versus the harvested energy level. This design was then extended to the broadcast channel with multiple ID receivers and multiple EH receivers (see, e.g., \cite{xu2014multiuser}). On the other hand, the works \cite{xiong2022flowing} and \cite{hua2022mimo} considered the basic ISAC setup with one BS, one ID receiver, and one sensing target, in which the transmit strategies at the BS were optimized to balance the communication rate versus the estimation Cram{\'e}r-Rao bound (CRB) as the sensing performance metric. Furthermore, the authors in \cite{liu2020joint} and \cite{hua2021optimal} studied the ISAC system with multiple ID receivers and sensing targets, in which the transmit beamforming design at the BS was optimized to balance the communication and sensing performances.

Different from these prior works, this paper studies a multi-functional MIMO system unifying ISAC and SWIPT as shown in Fig. 1, in which one single multi-antenna hybrid access point (H-AP) transmits wireless signals to simultaneously deliver information to one multi-antenna ID receiver, provide energy supply to one multi-antenna EH receiver, and estimate a sensing target based on the echo signals. For such a multi-functional wireless system, how to design the transmit strategies at the multi-antenna H-AP is essential to optimize the performance tradeoffs among sensing, communication, and powering. This problem, however, is particularly challenging, since the MIMO sensing (e.g., \cite{li2008range, bekkerman2006target}), communication (e.g., \cite{telatar1999capacity}), and powering (e.g., \cite{zhang2013mimo}) are designed based on distinct objectives and follow different principles. This thus motivates our study in this work.
\begin{figure}[tb]
	\centering
	{\includegraphics[width=0.32\textwidth]{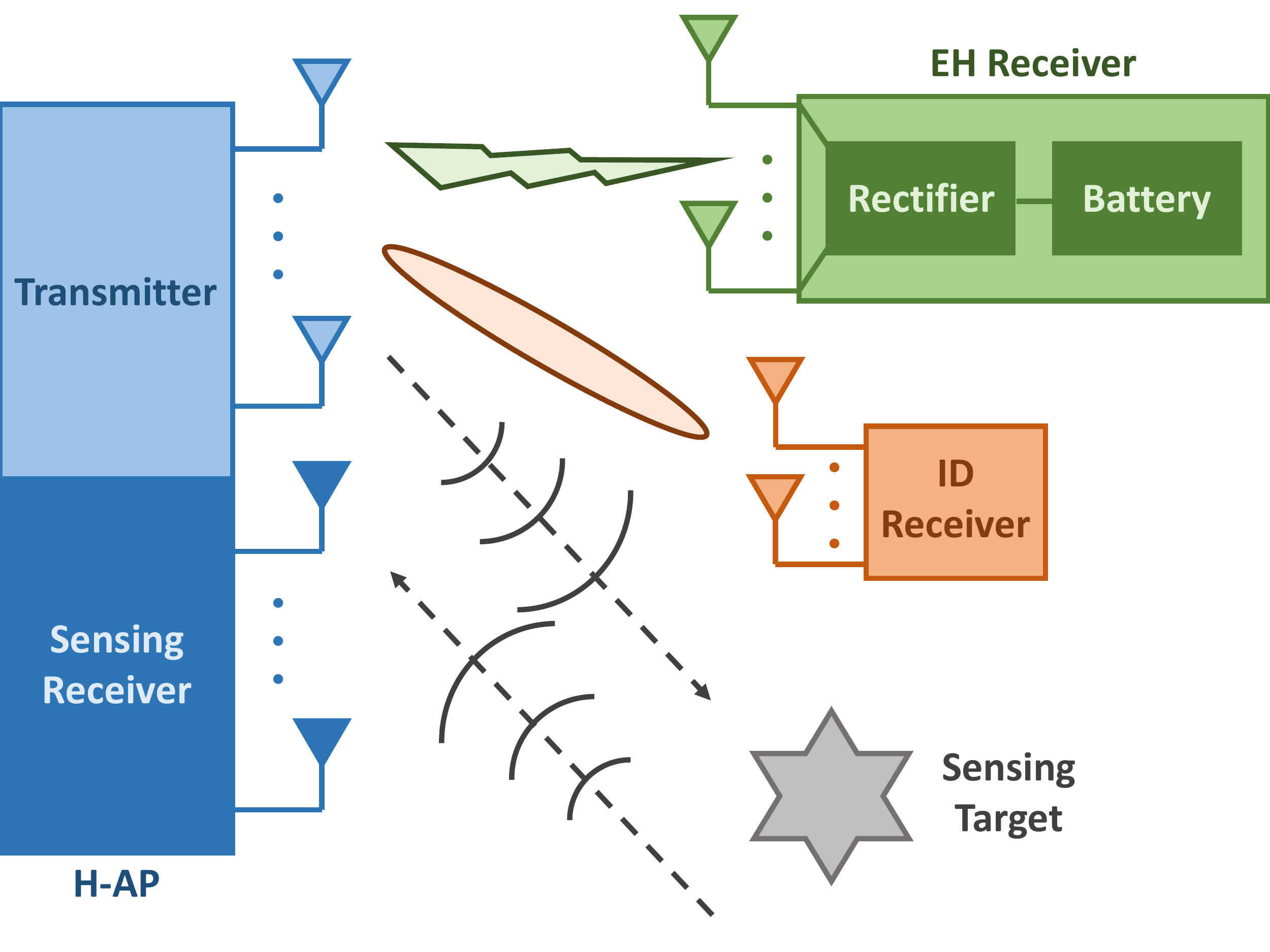}}
	\caption{A multi-functional MIMO system unifying SWIPT and ISAC.}
\end{figure}

In this paper, we consider the multi-functional MIMO system by focusing on the point target model for radar sensing, in which the H-AP aims to estimate the target angle. We aim to reveal the fundamental performance tradeoffs among sensing, communication, and powering, in terms of the angle estimation CRB, the achievable communication rate, and the harvested energy level. Towards this end, we characterize the complete CRB-rate-energy (C-R-E) region of this system, which is defined as the set of all C-R-E pairs that are simultaneously achievable. To find the points on the Pareto boundary surface of the C-R-E region, we propose to maximize the MIMO communication rate while ensuring the estimation CRB requirement for target sensing and the harvested energy requirement at the EH receiver, subject to the maximum transmit power constraint at the H-AP. We derive the well-structured optimal transmit covariance solution by using advanced convex optimization techniques. It is shown that the optimal transmit covariance follows the eigenmode transmission (EMT) structure based on a composite channel matrix, which can be generally divided into two parts, one for the triple roles of communication, sensing, and powering, and the other for sensing and powering only. Finally, numerical results validate the performance of our proposed design, in comparison to the benchmark scheme based on time switching.
 
\textit{Notations:} Boldface letters are used for vectors (lower-case) and matrices (upper-case). For a square matrix \(\boldsymbol{A}\), \(\mathrm{tr}(\boldsymbol{A})\) and \(\det(\boldsymbol{A})\) denote its trace and determinant, respectively, \(\boldsymbol{A} \succeq \boldsymbol{0}\) means that \(\boldsymbol{A}\) is positive semi-definite, and \(\boldsymbol{A} \succ \boldsymbol{0}\) means that \(\boldsymbol{A}\) is positive definite. For an arbitrary-size matrix \(\boldsymbol{A}\), \(\mathrm{rank}(\boldsymbol{A})\), \(\boldsymbol{A}^\dag\), \(\boldsymbol{A}^T\), \(\boldsymbol{A}^H\), and \(\mathcal{R}(\boldsymbol{A})\) denote its rank, conjugate, transpose, conjugate transpose, and range space, respectively. For a vector \(\boldsymbol{a}\), \(\|\boldsymbol{a}\|\) denotes its Euclidean norm. For a complex number \(a\), \(|a|\) denotes its magnitude. For a real number \(a\), \((a)^+ \triangleq \max(a,0)\). \(\mathrm{diag}(a_1, \dots, a_N)\) denotes a diagonal matrix whose diagonal entries are \(a_1, \dots, a_N\). \(\boldsymbol{I}_N\) denotes the identity matrix with dimension \(N \times N\). \(\mathbb{C}^{M \times N}\) and \(\mathbb{S}^{N}\)denote the spaces of \(M \times N\) complex matrices and \(N \times N\) Hermitian matrices. \(\mathbb{E}[\cdot]\) denotes the statistic expectation. \(j = \sqrt{-1}\).

\section{System Model}

This paper considers a multi-functional MIMO system as shown in Fig. 1, which consists of one H-AP, one EH receiver, one ID receiver, and one sensing target to be estimated. The H-AP is equipped with a uniform linear array (ULA) of \(M > 1\) transmit antennas and \(N_{\text{S}} > 1\) receive antennas for sensing. The EH and ID receivers are equipped with \(N_{\text{EH}} \ge 1\) and \(N_{\text{ID}} \ge 1\) receive antennas, respectively. 

We consider a quasi-static narrowband channel model, in which the wireless channels remain unchanged over the interested transmission block consisting of \(L\) symbols, where $L$ is assumed to be sufficiently large. Let \(\boldsymbol{H}_{\text{EH}} \in \mathbb{C}^{N_{\text{EH}} \times M}\) and \(\boldsymbol{H}_{\text{ID}} \in \mathbb{C}^{N_{\text{ID}} \times M}\) denote the channel matrices from the H-AP to the EH receiver and the ID receiver, respectively. Let \(\boldsymbol{H}_{\text{S}} \in \mathbb{C}^{N_{\text{S}} \times M}\) denote the target response matrix from the H-AP transmitter to the sensing target to the H-AP receiver, which will be specified later. It is assumed that the H-AP perfectly knows the information of \(\boldsymbol{H}_{\text{EH}}\) and \(\boldsymbol{H}_{\text{ID}}\), and the ID receiver perfectly knows \(\boldsymbol{H_{\text{ID}}}\) to facilitate the transmit optimization, as commonly assumed in the literature \cite{zhang2013mimo, xu2014multiuser}.

At each symbol \(l \in \{1, \dots, L\}\), let \(\boldsymbol{x}(l) \in \mathbb{C}^{M \times 1}\) denote the transmitted signal at the H-AP. The transmitted signal over the whole block is expressed as
\begin{equation}
	\boldsymbol{X} = \big[\boldsymbol{x}(1), \dots, \boldsymbol{x}(L)\big] \in \mathbb{C}^{M \times L}.
\end{equation}
Without loss of optimality, we consider the capacity-achieving Gaussian signaling, such that \(\boldsymbol{x}(l)\)'s are assumed to be circularly symmetric complex Gaussian (CSCG) random vectors with zero mean and covariance matrix  \(\boldsymbol{S} = \mathbb{E}\big[\boldsymbol{x}(l) \boldsymbol{x}(l)^H\big] \succeq \boldsymbol{0}\). As \(L\) is sufficiently large, the sample covariance matrix $\frac{1}{L} \boldsymbol{X} \boldsymbol{X}^H$ is approximated as the statistical covariance matrix $\boldsymbol{S}$, i.e., \(\frac{1}{L} \boldsymbol{X} \boldsymbol{X}^H \approx \boldsymbol{S}\), which is the optimization variable to be designed\footnote{This approximation has been widely adopted in MIMO radar and MIMO ISAC systems \cite{liu2021cramer, hua2022mimo}.}. Suppose that the H-AP is subject to a maximum transmit power budget \(P\). We thus have \(\mathbb{E}\big[\|\boldsymbol{x}(l)\|^2\big] = \mathrm{tr}(\boldsymbol{S}) \le P\).

First, we consider the radar sensing with a point target. Let \(\alpha\) denote the complex reflection coefficient, and \(\theta\) denote the angle of departure/arrival (AoD/AoA). The echo signal received by the H-AP receiver is
\begin{equation} \label{Y}
	\boldsymbol{Y}_{\text{S}} = \boldsymbol{H}_{\text{S}} \boldsymbol{X} + \boldsymbol{Z}_{\text{S}} = \alpha \boldsymbol{A}(\theta) \boldsymbol{X} + \boldsymbol{Z}_{\text{S}},
\end{equation}
where \(\boldsymbol{A}(\theta) \triangleq \boldsymbol{a}_r(\theta) \boldsymbol{a}_t^T(\theta)\), with \(\boldsymbol{a}_r(\theta) \in \mathbb{C}^{N_{\text{S}} \times 1}\) and \(\boldsymbol{a}_t(\theta) \in \mathbb{C}^{M \times 1}\) denoting the receive and transmit array steering vectors, and \(\boldsymbol{Z}_{\text{S}} \in \mathbb{C}^{N_{\text{S}} \times L}\) denotes the additive white Gaussian noise (AWGN) at the H-AP receiver that is a CSCG random matrix with independent and identically distributed (i.i.d.) entries each with zero mean and variance \(\sigma_{\text{S}}^2\). By choosing the center of the ULA antennas as the reference point and assuming half-wavelength spacing between adjacent antennas, we have
\begin{equation} \label{atar}
	\begin{aligned}
		\boldsymbol{a}_t(\theta) &= \big[e^{-j \frac{M-1}{2} \pi \sin\theta}, e^{-j \frac{M-3}{2} \pi \sin\theta}, \dots, e^{j \frac{M-1}{2} \pi \sin\theta}\big]^T, \\
		\boldsymbol{a}_r(\theta) &= \big[e^{-j \frac{N_{\text{S}}-1}{2} \pi \sin\theta}, e^{-j \frac{N_{\text{S}}-3}{2} \pi \sin\theta}, \dots, e^{j \frac{N_{\text{S}}-1}{2} \pi \sin\theta}\big]^T.
	\end{aligned}
\end{equation}
The objective of sensing is to estimate \(\theta\) and \(\alpha\) as the unknown parameters from the received echo signal \(\boldsymbol{Y}_{\text{S}}\) in \eqref{Y}, in which the transmitted signal \(\boldsymbol{X}\) is known by the H-AP. In this case, the CRB for estimating \(\theta\) is adopted as the sensing performance metric\footnote{As the target information contained in \(\alpha\) is hard to extract, in this paper we focus on the CRB for estimating \(\theta\), similarly as in prior works \cite{liu2021cramer, song2022intelligent}.}, which is given by \cite{bekkerman2006target}
\begin{equation} \label{CRB1}
	\mathrm{CRB}(\boldsymbol{S}) = \frac{\sigma_{\text{S}}^2}{2|\alpha|^2L} \frac{\mathrm{tr}(\! \boldsymbol{A}^H \! \boldsymbol{A} \boldsymbol{S})}{\mathrm{tr}(\! \boldsymbol{\dot{A}}^H \! \boldsymbol{\dot{A}} \boldsymbol{S}) \mathrm{tr}(\! \boldsymbol{A}^H \! \boldsymbol{A} \boldsymbol{S}) - |\mathrm{tr}(\! \boldsymbol{\dot{A}}^H \! \boldsymbol{A} \boldsymbol{S})|^2},   
\end{equation}
where we define \(\boldsymbol{A} \triangleq \boldsymbol{A}(\theta)\) and \(\boldsymbol{\dot{A}} \triangleq \frac{\partial \boldsymbol{A}(\theta)}{\partial \theta}\).

Next, we consider the point-to-point MIMO communication from the H-AP to the ID receiver. The received signal by the ID receiver at symbol \(l\) is given by
\begin{equation}
	\boldsymbol{y}_{\text{ID}}(l) = \boldsymbol{H}_{\text{ID}} \boldsymbol{x}(l) + \boldsymbol{z}_{\text{ID}}(l),
\end{equation}
where \(\boldsymbol{z}_{\text{ID}}(l) \in \mathbb{C}^{N_{\text{ID}} \times 1}\) denotes the AWGN at the ID receiver that is a CSCG random vector with zero mean and covariance matrix \(\sigma_{\text{ID}}^2 \boldsymbol{I}_{N_{\text{ID}}}\). With the capacity-achieving Gaussian signaling, the achievable data rate (in bps/Hz) is \cite{telatar1999capacity}
\begin{equation} \label{Rate}
	{R}(\boldsymbol{S}) = \log_2 \det \Big(\boldsymbol{I}_{N_{\text{ID}}} + \frac{1}{\sigma_{\text{ID}}^2} \boldsymbol{H}_{\text{ID}} \boldsymbol{S} \boldsymbol{H}_{\text{ID}}^H\Big).
\end{equation}

Then, we consider the WPT from the H-AP to the EH receiver, where the EH receiver uses the rectifiers to convert the received radio frequency (RF) signals into direct current (DC) signals for energy harvesting. In general, the harvested DC power is monotonically increasing with respect to the received RF power. As a result, we use the received RF power (energy over a unit time period, in Watt) at the EH receiver as the powering performance metric \cite{clerckx2019fundamentals}, i.e.,
\begin{equation} \label{E}
	{E}(\boldsymbol{S}) = \mathrm{tr}(\boldsymbol{H}_{\text{EH}} \boldsymbol{S} \boldsymbol{H}_{\text{EH}}^H).
\end{equation}

Our objective is to reveal the fundamental performance tradeoff among the estimation CRB \(\mathrm{CRB}(\boldsymbol{S})\) in \eqref{CRB1}, the achievable rate \({R}(\boldsymbol{S})\) in \eqref{Rate}, and the received (unit-time) energy \({E}(\boldsymbol{S})\) in \eqref{E}. Towards this end, we characterize the C-R-E region that is defined as all the C-R-E pairs simultaneously achievable in the multi-functional wireless MIMO system, i.e.,
\begin{equation} \label{CRE}
	\begin{aligned}
		\mathcal{C} \triangleq \big\{(\widehat{\mathrm{CRB}}, \widehat{{R}}, \widehat{{E}}) | & \widehat{\mathrm{CRB}} \ge \mathrm{CRB}(\boldsymbol{S}), \widehat{{R}} \le {R}(\boldsymbol{S}), \\
		& \widehat{{E}} \le {E}(\boldsymbol{S}), \mathrm{tr}(\boldsymbol{S}) \le P, \boldsymbol{S} \succeq \boldsymbol{0}\big\}.
	\end{aligned}
\end{equation}

\section{C-R-E Region Characterization}

This section characterizes the C-R-E region of the multi-functional MIMO system \(\mathcal{C}\) in \eqref{CRE}, by finding its Pareto boundary, which corresponds to the set of points at which improving one performance metric will inevitably result in the deterioration of another. The complete Pareto boundary corresponds to a surface in a three-dimensional (3D) space as shown in Fig. 2, which is surrounded by three vertices corresponding to rate maximization (R-max), energy maximization (E-max), and CRB minimization (C-min), respectively, as well as three edges corresponding to the optimal C-R, R-E, and C-E tradeoffs, respectively.
\begin{figure}[tb]
	\centering
	{\includegraphics[width=0.32\textwidth]{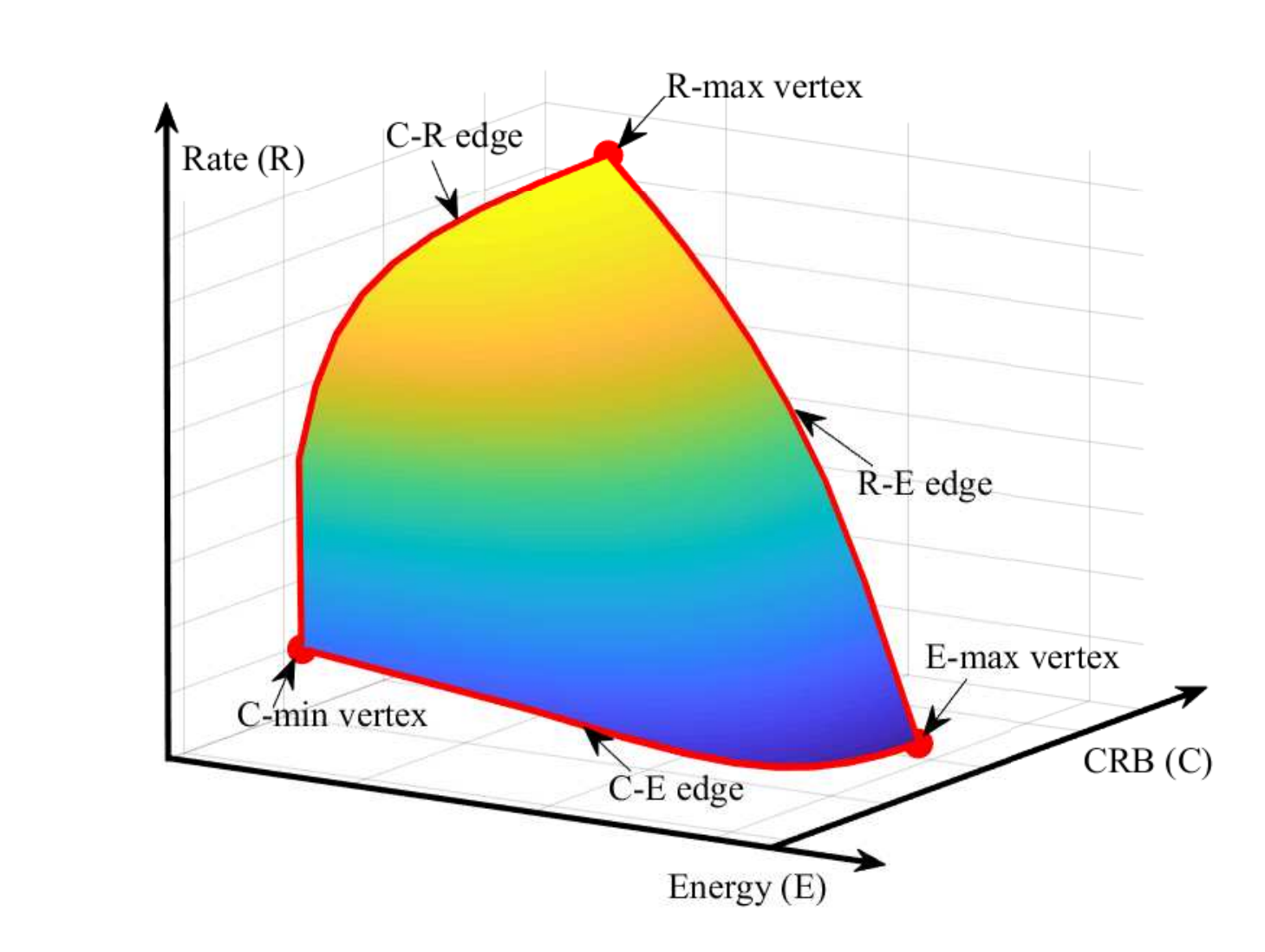}}
	\caption{The Pareto boundary of an example C-R-E region.}
\end{figure}

\subsection{Finding Three Vertices of Pareto Boundary}\label{Section:III-A}

This subsection finds the three vertices of the C-R-E region \(\mathcal{C}\). First, we find the R-max vertex \((\mathrm{CRB}_{\text{ID}}, {R}_{\mathrm{max}}, {E}_{\text{ID}})\) by optimizing the transmit covariance $\boldsymbol{S}$ to maximize the communication rate ${R}(\boldsymbol{S})$, i.e., \(\max_{\boldsymbol{S} \succeq \boldsymbol{0}} {R}(\boldsymbol{S}), \ \mathrm{s.t.} \ \mathrm{tr}(\boldsymbol{S}) \le P\). This corresponds to a sole MIMO communication system. It has been established in \cite{telatar1999capacity} that the optimal transmit covariance solution to this problem, denoted by \(\boldsymbol{S}_{\text{ID}}\), can be obtained by the EMT over the ID channel (i.e., performing the singular value decomposition (SVD) to decompose $\boldsymbol{H}_{\text{ID}}$) together with the water-filling power allocation. Accordingly, the maximum communication rate is \({R}_{\mathrm{max}} = {R}(\boldsymbol{S}_{\text{ID}})\), and the corresponding harvested energy level and estimation CRB are \({E}_{\text{ID}} = {E}(\boldsymbol{S}_{\text{ID}})\) and \(\mathrm{CRB}_{\text{ID}} = \mathrm{CRB}(\boldsymbol{S}_{\text{ID}})\), respectively.

Next, we consider the E-max vertex \((\mathrm{CRB}_{\text{EH}}, {R}_{\text{EH}}, {E}_{\mathrm{max}})\), which can be found by optimizing $\boldsymbol{S}$ to maximize the harvested energy level ${E}(\boldsymbol{S})$, i.e., \(\max_{\boldsymbol{S} \succeq \boldsymbol{0}} {E}(\boldsymbol{S}), \ \mathrm{s.t.} \ \mathrm{tr}(\boldsymbol{S}) \le P\). This corresponds to a sole MIMO WPT system. It has been shown in \cite{zhang2013mimo} that the optimal solution is a rank-one matrix that can be obtained based on the strongest eigenmode transmission, denoted by \(\boldsymbol{S}_{\text{EH}} = \sqrt{P}\boldsymbol{v}_{\max} \boldsymbol{v}_{\max}^H\), with $\boldsymbol{v}_{\max}$ being the dominant right singular vector of $\boldsymbol{H}_{\mathrm{EH}}$. The maximum harvested energy level is hence \({E}_{\mathrm{max}} = {E}(\boldsymbol{S}_{\text{EH}})\), and the corresponding communication rate and estimation CRB are \({R}_{\text{EH}} = {R}(\boldsymbol{S}_{\text{EH}})\) and \(\mathrm{CRB}_{\text{EH}} = \mathrm{CRB}(\boldsymbol{S}_{\text{EH}})\), respectively.

Finally, we consider the C-min vertex \((\mathrm{CRB}_{\mathrm{min}}, {R}_{\text{S}}, {E}_{\text{S}})\), which can be found by optimizing $\boldsymbol{S}$ to minimize the estimation CRB $\mathrm{CRB}(\boldsymbol{S})$, i.e., \(\min_{\boldsymbol{S} \succeq \boldsymbol{0}} \mathrm{CRB}(\boldsymbol{S}), \ \mathrm{s.t.} \ \mathrm{tr}(\boldsymbol{S}) \le P\). This corresponds to a sole MIMO radar sensing system. The optimal solution \(\boldsymbol{S}_{\text{S}}\) has been derived in closed form in \cite{li2008range}, where \(\boldsymbol{S}_{\text{S}} = \frac{P}{M}\boldsymbol{a}_t(\theta)^\dag \boldsymbol{a}_t(\theta)^T\) holds if \(N_{\text{S}} > M\). Thus, the minimum CRB is \(\mathrm{CRB}_{\mathrm{min}} = \mathrm{CRB}(\boldsymbol{S}_{\text{S}})\), and the corresponding communication rate and harvested energy level are \({R}_{\text{S}} = {R}(\boldsymbol{S}_{\text{S}})\) and \({E}_{\text{S}} = {E}(\boldsymbol{S}_{\text{S}})\), respectively.

\subsection{Finding Three Edges of Pareto Boundary}\label{Section:III-B}

This subsection finds the three edges on the Pareto boundary of the C-R-E region \(\mathcal{C}\). First, we find the C-R edge connecting the C-min vertex \((\mathrm{CRB}_{\mathrm{min}}, {R}_{\text{S}}, {E}_{\text{S}})\) and the R-max vertex \((\mathrm{CRB}_{\text{ID}}, {R}_{\mathrm{max}}, {E}_{\text{ID}})\), by optimizing $\boldsymbol{S}$ to maximize the communication rate ${R}(\boldsymbol{S})$ subject to a maximum estimation CRB constraint, i.e., \(\max_{\boldsymbol{S} \succeq \boldsymbol{0}} {R}(\boldsymbol{S}), \mathrm{s.t.} \mathrm{CRB}(\boldsymbol{S}) \le \mathrm{CRB}_{\text{C-R}}, \mathrm{tr}(\boldsymbol{S}) \le P\), where the CRB threshold \(\mathrm{CRB}_{\text{C-R}}\) ranges from \(\mathrm{CRB}_{\mathrm{min}}\) to \(\mathrm{CRB}_{\text{ID}}\) in order to get different C-R tradeoff points on the edge. This corresponds to the CRB-constrained rate maximization problem for the MIMO ISAC system without WPT, for which the optimal solution has been derived in  \cite{hua2022mimo} in semi-closed-form, denoted by \(\boldsymbol{S}_{\text{C-R}}\).

Next, we find the R-E edge connecting the R-max vertex \((\mathrm{CRB}_{\text{ID}}, {R}_{\mathrm{max}}, {E}_{\text{ID}})\) and the E-max vertex \((\mathrm{CRB}_{\text{EH}}, {R}_{\text{EH}}, {E}_{\mathrm{max}})\). Towards this end, we optimize $\boldsymbol{S}$ to maximize the communication rate ${R}(\boldsymbol{S})$ subject to a minimum harvested energy constraint, i.e., \(\max_{\boldsymbol{S} \succeq \boldsymbol{0}} {R}(\boldsymbol{S}), \mathrm{s.t.} {E}(\boldsymbol{S}) \ge {E}_{\text{R-E}}, \mathrm{tr}(\boldsymbol{S}) \le P\), where the energy threshold \({E}_{\text{R-E}}\) ranges from \({E}_{\text{ID}}\) to \({E}_{\mathrm{max}}\) in order to get different R-E tradeoff points on the edge. This corresponds to the energy-constrained rate maximization problem for the MIMO SWIPT system without radar sensing, whose optimal transmit covariance solution has been obtained in \cite{zhang2013mimo} as \(\boldsymbol{S}_{\text{R-E}}\) in semi-closed-form.

Then, we find the C-E edge to connect the C-min vertex \((\mathrm{CRB}_{\mathrm{min}}, {R}_{\text{S}}, {E}_{\text{S}})\) and the E-max vertex \((\mathrm{CRB}_{\text{EH}}, {R}_{\text{EH}}, {E}_{\mathrm{max}})\). To achieve this, we optimize $\boldsymbol{S}$ to maximize the harvested energy level ${E}(\boldsymbol{S})$ subject to a maximum estimation CRB constraint, i.e., \(\max_{\boldsymbol{S} \succeq \boldsymbol{0}} {E}(\boldsymbol{S}), \mathrm{s.t.} \mathrm{CRB}(\boldsymbol{S}) \le \mathrm{CRB}_{\text{C-E}}, \mathrm{tr}(\boldsymbol{S}) \le P\), where the CRB threshold \(\mathrm{CRB}_{\text{C-E}}\) ranges from \(\mathrm{CRB}_{\mathrm{min}}\) to \(\mathrm{CRB}_{\text{EH}}\) in order to get different C-E tradeoff points on the edge. This corresponds to the case when there are only sensing and WPT in our integrated system. The optimization problem can be reformulated into a convex form, as the CRB constraint \(\mathrm{CRB}(\boldsymbol{S}) \le \mathrm{CRB}_{\text{C-E}}\) is equivalent to the convex semi-definite constraint \(\begin{bmatrix}
	\mathrm{tr}(\boldsymbol{\dot{A}}^H \boldsymbol{\dot{A}} \boldsymbol{S}) - \frac{\sigma_{\text{S}}^2}{2 |\alpha|^2 L \mathrm{CRB}_{\text{C-E}}} & \mathrm{tr}(\boldsymbol{\dot{A}}^H \boldsymbol{A} \boldsymbol{S})^\dag \\
	\mathrm{tr}(\boldsymbol{\dot{A}}^H \boldsymbol{A} \boldsymbol{S}) & \mathrm{tr}(\boldsymbol{A}^H \boldsymbol{A} \boldsymbol{S})
\end{bmatrix} \succeq \boldsymbol{0}\) based on the Schur complement \cite{liu2021cramer}. Therefore, the optimal solution \(\boldsymbol{S}_{\text{C-E}}\) in this case can be found by standard convex optimization techniques\footnote{Note that we can obtain a well-structured optimal solution to this problem by the Lagrangian duality method. However, we omit the derivation here as it will be similar to that for the more general problem (P1) in Section IV.}.

\subsection{Characterizing Complete Pareto Boundary Surface}\label{Section:III-C}

With the vertices and edges obtained, it remains to find the interior points on the Pareto boundary surface to characterize the complete C-R-E region \(\mathcal{C}\). Towards this end, we optimize the transmit covariance $\boldsymbol{S}$ to maximize the communication rate ${R}(\boldsymbol{S})$ in \eqref{Rate}, subject to a maximum estimation CRB constraint $\Gamma_{\text{S}}$ and a minimum harvested energy constraint $\Gamma_{\text{EH}}$. Mathematically, the CRB-and-energy-constrained rate maximization problem is formulated as 
\begin{subequations}
\begin{align} 
	(\text{P1}): \max_{\boldsymbol{S} \succeq \boldsymbol{0}} &\ \log_2 \det \Big(\boldsymbol{I}_{N_{\text{ID}}} + \frac{1}{\sigma_{\text{ID}}^2} \boldsymbol{H}_{\text{ID}} \boldsymbol{S} \boldsymbol{H}_{\text{ID}}^H\Big) \label{R} \\
	\mathrm{s.t.} &\ \mathrm{tr}(\boldsymbol{H}_{\text{EH}} \boldsymbol{S} \boldsymbol{H}_{\text{EH}}^H) \ge \Gamma_{\text{EH}} \label{Ra} \\
	&\ \mathrm{CRB}(\boldsymbol{S}) \le \Gamma_{\text{S}} \label{Rb} \\ 
	&\ \mathrm{tr}(\boldsymbol{S}) \le P. \label{Rc}
\end{align}
\end{subequations}
By exhausting \(\Gamma_{\text{EH}}\) and \(\Gamma_{\text{S}}\) enclosed by the projection of the three edges on the C-E plane, we can obtain all the Pareto boundary surface points. In particular, let \(R^\star(\Gamma_{\text{EH}}, \Gamma_{\text{S}})\) denote the optimal value of problem (P1) with given \(\Gamma_{\text{EH}}\) and \(\Gamma_{\text{S}}\). Then we have one Pareto boundary point of the C-R-E region \(\mathcal{C}\) as \(\big(\Gamma_{\text{S}}, R^\star(\Gamma_{\text{EH}}, \Gamma_{\text{S}}), \Gamma_{\text{EH}}\big)\). 

\section{Optimal Solution to Problem (P1)}

In this section, we present the optimal solution to problem (P1). Notice that (P1) is not convex due to the estimation CRB constraint in \eqref{Rb}. Fortunately, based on the Schur complement, constraint \eqref{Rb} is equivalent to
\begin{equation}\label{Rb-reformulation}
\begin{bmatrix}
	\mathrm{tr}(\boldsymbol{\dot{A}}^H \boldsymbol{\dot{A}} \boldsymbol{S}) - \frac{1}{\Gamma_{\text{S},1}} & \mathrm{tr}(\boldsymbol{\dot{A}}^H \boldsymbol{A} \boldsymbol{S})^\dag \\
	\mathrm{tr}(\boldsymbol{\dot{A}}^H \boldsymbol{A} \boldsymbol{S}) & \mathrm{tr}(\boldsymbol{A}^H \boldsymbol{A} \boldsymbol{S})
\end{bmatrix} \succeq \boldsymbol{0},
\end{equation}
where \(\Gamma_{\text{S},1} \triangleq \frac{2|\alpha|^2L}{\sigma_{\text{S}}^2} \Gamma_{\text{S}}\). Thus, by replacing constraint \eqref{Rb} as that in \eqref{Rb-reformulation}, (P1) is reformulated in a convex form, and can be optimally solved by using the Lagrangian duality method. 

Let \(\lambda \ge 0\), \(\nu \ge 0\), and \(\boldsymbol{Z} \triangleq \begin{bmatrix}
	z_1 & z_2 \\ z_2^\dag & z_3
\end{bmatrix} \succeq \boldsymbol{0}\) denote the dual variables associated with the constraints in \eqref{Ra}, \eqref{Rc}, and \eqref{Rb-reformulation}, respectively. The Lagrangian of problem (P1) is 
\begin{equation}
	\begin{aligned}
		\mathcal{L}(\boldsymbol{S}, \lambda, \nu, \boldsymbol{Z}) =& \log_2 \det \Big(\boldsymbol{I}_{N_{\text{ID}}} + \frac{1}{\sigma_{\text{ID}}^2} \boldsymbol{H}_{\text{ID}} \boldsymbol{S} \boldsymbol{H}_{\text{ID}}^H\Big) \\
		-& \mathrm{tr}(\boldsymbol{D} \boldsymbol{S}) - \lambda \Gamma_{\text{EH}} + \nu P - \frac{z_1}{\Gamma_{\text{S},1}},
	\end{aligned}
\end{equation}
where \(\boldsymbol{D} \triangleq \nu \boldsymbol{I}_M - \lambda \boldsymbol{H}_{\text{EH}}^H \boldsymbol{H}_{\text{EH}} -  z_1 \boldsymbol{\dot{A}}^H \boldsymbol{\dot{A}} - z_2 \boldsymbol{\dot{A}}^H \boldsymbol{A} - z_2^\dag \boldsymbol{A}^H \boldsymbol{\dot{A}} - z_3 \boldsymbol{A}^H \boldsymbol{A}\). 

Accordingly, the dual function of (P1) is defined as 
\begin{equation} \label{dualfunc_p}
	g(\lambda, \nu, \boldsymbol{Z}) = \max_{\boldsymbol{S} \succeq \boldsymbol{0}} \ \mathcal{L}(\boldsymbol{S},\lambda, \nu, \boldsymbol{Z}).
\end{equation}
We have the following lemma, which can be verified similarly as in \cite[Lemma 1]{hua2022mimo}. 

\begin{lemma} \label{lem1}
	In order for the dual function \(g(\lambda, \nu, \boldsymbol{Z})\) to be bounded from above, it must hold that 
	\begin{equation}\label{D1-constraint}
		\boldsymbol{D} \succeq \boldsymbol{0}~{\text{and}}~\mathcal{R}(\boldsymbol{H}_{\text{ID}}^H) \subseteq \mathcal{R}(\boldsymbol{D}).
	\end{equation}
\end{lemma}

Based on Lemma \ref{lem1}, the dual problem of (P1) is defined as 
\begin{equation}
	(\text{D}1): \min_{\lambda \ge 0, \nu \ge 0, \boldsymbol{Z} \succeq \boldsymbol{0}} \ g(\lambda, \nu, \boldsymbol{Z}), \ \mathrm{s.t.} \ \eqref{D1-constraint}.
\end{equation}

Since problem (P1) is reformulated in a convex form and satisfies Slater's condition, strong duality holds between (P1) and its dual problem (D1) \cite{boyd2004vandenberghe}. Therefore, we can solve (P1) by equivalently solving (D1). In the following, we first obtain the dual function \(g(\lambda, \nu, \boldsymbol{Z})\) with given dual variables, then search over \(\lambda \ge 0\), \(\nu \ge 0\), and \(\boldsymbol{Z} \succeq \boldsymbol{0}\) to minimize \(g(\lambda, \nu, \boldsymbol{Z})\).

\subsection{Finding Dual Function \(g(\lambda, \nu, \boldsymbol{Z})\)}

First, we find the dual function \(g(\lambda, \nu, \boldsymbol{Z})\) under given \(\lambda \ge 0\), \(\nu \ge 0\), and \(\boldsymbol{Z} \succeq \boldsymbol{0}\) satisfying \eqref{D1-constraint}. By dropping the constant terms \(-\lambda \Gamma_{\text{EH}} + \nu P - \frac{z_1}{\Gamma_{\text{S},1}}\), problem \eqref{dualfunc_p} is equivalent to
\begin{equation} \label{L_P_1}
	\max_{\boldsymbol{S} \succeq \boldsymbol{0}} \ \log_2 \det \Big(\boldsymbol{I}_{N_{\text{ID}}} + \frac{1}{\sigma_{\text{ID}}^2} \boldsymbol{H}_{\text{ID}} \boldsymbol{S} \boldsymbol{H}_{\text{ID}}^H\Big) - \mathrm{tr}(\boldsymbol{D} \boldsymbol{S}).
\end{equation}
Suppose that \(\mathrm{rank}(\boldsymbol{D}) = r_{\text{p}}\), the eigenvalue decomposition (EVD) of \(\boldsymbol{D}\) is expressed as
\begin{equation} \label{D}
	\boldsymbol{D} = 
	\begin{bmatrix}
		\boldsymbol{Q}^{\overline{\mathrm{null}}} \ \boldsymbol{Q}^{\mathrm{null}}
	\end{bmatrix}
	\begin{bmatrix}
		\boldsymbol{\Sigma}^{\overline{\mathrm{null}}} & \boldsymbol{0} \\ \boldsymbol{0} & \boldsymbol{0}
	\end{bmatrix}
	\begin{bmatrix}
		{\boldsymbol{Q}^{\overline{\mathrm{null}}}} \ {\boldsymbol{Q}^{\mathrm{null}}}
	\end{bmatrix}^H,
\end{equation}
where \(\boldsymbol{\Sigma}^{\overline{\mathrm{null}}} = \mathrm{diag}(\sigma_{\text{p},1}, \dots, \sigma_{\text{p},r_{\text{p}}})\), and \({\boldsymbol{Q}^{\overline{\mathrm{null}}}} \in \mathbb{C}^{M \times r_{\text{p}}}\) and \({\boldsymbol{Q}^{\mathrm{null}}} \in \mathbb{C}^{M \times (M-r_{\text{p}})}\) denote the eigenvectors corresponding to the \(r_{\text{p}}\) non-zero eigenvalues \(\sigma_{\text{p},1}, \dots, \sigma_{\text{p},r_{\text{p}}}\), and the remaining \(M-r_{\text{p}}\) zero eigenvalues, respectively. Without loss of generality, we express \(\boldsymbol{S}\) as
\begin{equation} \label{S_p}
	\boldsymbol{S} = 
	\begin{bmatrix}
		\boldsymbol{Q}^{\overline{\mathrm{null}}} \ \boldsymbol{Q}^{\mathrm{null}}
	\end{bmatrix}
	\begin{bmatrix}
		\boldsymbol{S}_{11} & \boldsymbol{S}_{10} \\ \boldsymbol{S}_{10}^H & \boldsymbol{S}_{00}
	\end{bmatrix}
	\begin{bmatrix}
			{\boldsymbol{Q}^{\overline{\mathrm{null}}}} \ {\boldsymbol{Q}^{\mathrm{null}}}
	\end{bmatrix}^H,
\end{equation}
where \(\boldsymbol{S}_{11} \in \mathbb{S}^{r_{\text{p}}}\), \(\boldsymbol{S}_{00} \in \mathbb{S}^{M-r_{\text{p}}}\), and \(\boldsymbol{S}_{10} \in \mathbb{C}^{r_{\text{p}} \times (M-r_{\text{p}})}\) are variables to be optimized.

According to Lemma \ref{lem1}, we only need to deal with problem \eqref{L_P_1} for the case with \(\boldsymbol{H}_{\text{ID}} (\boldsymbol{Q}^{\overline{\mathrm{null}}} \boldsymbol{S}_{10} {\boldsymbol{Q}^{\mathrm{null}}}^H + \boldsymbol{Q}^{\mathrm{null}} \boldsymbol{S}_{10}^H {\boldsymbol{Q}^{\overline{\mathrm{null}}}}^H+ \boldsymbol{Q}^{\mathrm{null}} \boldsymbol{S}_{00} {\boldsymbol{Q}^{\mathrm{null}}}^H) \boldsymbol{H}_{\text{ID}}^H = \boldsymbol{0}\). Therefore, \eqref{L_P_1} can be simplified as the optimization of \(\boldsymbol{S}_{11}\) in the following.
\begin{equation} \label{L_P_111}
	\begin{aligned}
		\max_{\boldsymbol{S}_{11} \succeq \boldsymbol{0}} \ &\log_2 \det \Big(\boldsymbol{I}_{N_{\text{ID}}} + \frac{1}{\sigma_{\text{ID}}^2} \boldsymbol{H}_{\text{ID}} \boldsymbol{Q}^{\overline{\mathrm{null}}} \boldsymbol{S}_{11} {\boldsymbol{Q}^{\overline{\mathrm{null}}}}^H \boldsymbol{H}_{\text{ID}}^H\Big) \\
		&- \mathrm{tr}(\boldsymbol{\Sigma}^{\overline{\mathrm{null}}} \boldsymbol{S}_{11}).
	\end{aligned}
\end{equation}
Suppose that \(\mathrm{rank}\big(\boldsymbol{H}_{\text{ID}} \boldsymbol{Q}^{\overline{\mathrm{null}}} (\boldsymbol{\Sigma}^{\overline{\mathrm{null}}})^{-\frac{1}{2}}\big) = \widetilde{r}_{\text{p}}\). Then we have the SVD as \(\boldsymbol{H}_{\text{ID}} \boldsymbol{Q}^{\overline{\mathrm{null}}} (\boldsymbol{\Sigma}^{\overline{\mathrm{null}}})^{-\frac{1}{2}} = \boldsymbol{U}_{\text{p}} \boldsymbol{\Lambda}_{\text{p}} \boldsymbol{V}_{\text{p}}^H\), where \(\boldsymbol{U}_{\text{p}} \in \mathbb{C}^{N_{\text{ID}} \times N_{\text{ID}}}\), \(\boldsymbol{\Lambda}_{\text{p}}^H \boldsymbol{\Lambda}_{\text{p}} = \mathrm{diag}(\lambda_{\text{p},1}^2, \dots, \lambda_{\text{p},\widetilde{r}_{\text{p}}}^2, 0, \dots, 0)\), and \(\boldsymbol{V}_{\text{p}} \in \mathbb{C}^{r_{\text{p}} \times r_{\text{p}}}\). Substituting \(\boldsymbol{S}_{11} = (\boldsymbol{\Sigma}^{\overline{\mathrm{null}}})^{-\frac{1}{2}} \boldsymbol{V}_{\text{p}} \boldsymbol{\widetilde{S}}_{11} \boldsymbol{V}_{\text{p}}^H (\boldsymbol{\Sigma}^{\overline{\mathrm{null}}})^{-\frac{1}{2}}\), \eqref{L_P_111} is re-expressed as
\begin{equation} \label{L_P_2}
	\max_{\boldsymbol{\widetilde{S}}_{11} \succeq \boldsymbol{0}} \ \log_2 \det \Big(\boldsymbol{I}_{r_{\text{p}}} + \frac{1}{\sigma_{\text{ID}}^2} \boldsymbol{\Lambda}_{\text{p}}^H \boldsymbol{\Lambda}_{\text{p}} \boldsymbol{\widetilde{S}}_{11}\Big) - \mathrm{tr}(\boldsymbol{\widetilde{S}}_{11}).
\end{equation}
As shown in \cite{zhang2013mimo}, the optimal solution to problem \eqref{L_P_2} is given by \(\boldsymbol{\widetilde{S}}_{11}^* = \mathrm{diag}(\widetilde{p}_1, \dots, \widetilde{p}_{\widetilde{r}_{\text{p}}}, 0, \dots, 0)\), where
\begin{equation} \label{power_allo}
	\widetilde{p}_k = (\frac{1}{\ln 2}-\frac{\sigma_{\text{ID}}^2}{\lambda_{\text{p},k}^2})^+, \forall k \in \{1,\dots,\widetilde{r}_{\text{p}}\}.
\end{equation}

As a result, the optimal solution to problem \eqref{L_P_1} is given by \(\boldsymbol{S}^* = \begin{bmatrix}
	\boldsymbol{Q}^{\overline{\mathrm{null}}} \ \boldsymbol{Q}^{\mathrm{null}}
\end{bmatrix} \begin{bmatrix}
	\boldsymbol{S}_{11}^* & \boldsymbol{S}_{10}^* \\ {\boldsymbol{S}_{10}^*}^H & \boldsymbol{S}_{00}^*
\end{bmatrix} \begin{bmatrix}
\boldsymbol{Q}^{\overline{\mathrm{null}}} \ \boldsymbol{Q}^{\mathrm{null}}
\end{bmatrix}^H\), where
\begin{equation} \label{S_11}
	\boldsymbol{S}_{11}^* = (\boldsymbol{\Sigma}^{\overline{\mathrm{null}}})^{-\frac{1}{2}} \boldsymbol{V}_{\text{p}} \boldsymbol{\widetilde{S}}_{11}^* \boldsymbol{V}_{\text{p}}^H (\boldsymbol{\Sigma}^{\overline{\mathrm{null}}})^{-\frac{1}{2}},
\end{equation}
and \(\boldsymbol{S}_{10}^*\) and \(\boldsymbol{S}_{00}^* \succeq \boldsymbol{0}\) can be chosen arbitrarily such that \(\boldsymbol{S}^* \succeq \boldsymbol{0}\)\footnote{Note that \(\boldsymbol{S}_{10}^*\) and \(\boldsymbol{S}_{00}^*\) are non-unique here, and as a result, we need an additional step to determine them for solving the primal problem (P1) later. Here, we can simply choose \(\boldsymbol{S}_{10}^* = \boldsymbol{0}\) and \(\boldsymbol{S}_{00}^* = \boldsymbol{0}\) for obtaining the dual function \(g(\lambda,\nu,\boldsymbol{Z})\) only.}.

\subsection{Solving Dual Problem (D1)}

Next, we solve dual problem (D1), which is convex but not necessarily differentiable. Therefore, we can solve (D1) by applying subgradient-based methods such as the ellipsoid method \cite{boyd2004vandenberghe}. First, for the objective function \(g(\lambda, \nu, \boldsymbol{Z})\), the subgradient at \((\lambda, \nu, z_1, z_2, z_3)\) is \(\big[\mathrm{tr}(\boldsymbol{H}_{\text{EH}}^H \boldsymbol{H}_{\text{EH}} \boldsymbol{S}^*) - \Gamma_{\text{EH}},
P - \mathrm{tr}(\boldsymbol{S}^*),
\mathrm{tr}(\boldsymbol{\dot{A}}^H \boldsymbol{\dot{A}} \boldsymbol{S}^*) - \frac{1}{\Gamma_{\text{S},1}},
\mathrm{tr}(\boldsymbol{\dot{A}}^H \boldsymbol{A} \boldsymbol{S}^* + \boldsymbol{A}^H \boldsymbol{\dot{A}} \boldsymbol{S}^*) + j \mathrm{tr}(\boldsymbol{\dot{A}}^H \boldsymbol{A} \boldsymbol{S}^* - \boldsymbol{A}^H \boldsymbol{\dot{A}} \boldsymbol{S}^*),
\mathrm{tr}(\boldsymbol{A}^H \boldsymbol{A} \boldsymbol{S}^*)\big]^T\).
Then, let \(\boldsymbol{q}_1\) denote the eigenvector corresponding to the minimum eigenvalue of \(\boldsymbol{D}\). Since constraint \(\boldsymbol{D} \succeq \boldsymbol{0}\) is equivalent to \(\boldsymbol{q}_1^H \boldsymbol{D} \boldsymbol{q}_1 \ge 0\), the subgradient of constraint \(\boldsymbol{D} \succeq \boldsymbol{0}\) at \((\lambda, \nu, z_1, z_2, z_3)\) is \(\big[\boldsymbol{q}_1^H \boldsymbol{H}_{\text{EH}}^H \boldsymbol{H}_{\text{EH}} \boldsymbol{q}_1,
-1,
\boldsymbol{q}_1^H \boldsymbol{\dot{A}}^H \boldsymbol{\dot{A}} \boldsymbol{q}_1,
\boldsymbol{q}_1^H (\boldsymbol{\dot{A}}^H \boldsymbol{A} + \boldsymbol{A}^H \boldsymbol{\dot{A}}) \boldsymbol{q}_1 + j \boldsymbol{q}_1^H (\boldsymbol{\dot{A}}^H \boldsymbol{A} - \boldsymbol{A}^H \boldsymbol{\dot{A}}) \boldsymbol{q}_1,
\boldsymbol{q}_1^H \boldsymbol{A}^H \boldsymbol{A} \boldsymbol{q}_1\big]^T\).
Furthermore, let \(\boldsymbol{q}_2 = [q_{2,1}, q_{2,2}]^T\) denote the eigenvector corresponding to the minimum eigenvalue of \(\boldsymbol{Z}\). The subgradient of constraint \(\boldsymbol{Z} \succeq \boldsymbol{0}\) at \((\lambda, \nu, z_1, z_2, z_3)\) is \(\big[0,
0,
-|q_{2,1}|^2,
-(q_{2,1}^\dag q_{2,2} + q_{2,2}^\dag q_{2,1}) - j (q_{2,1}^\dag q_{2,2} - q_{2,2}^\dag q_{2,1}),
-|q_{2,2}|^2\big]^T\). With these derived subgradients, the ellipsoid method can be implemented efficiently, based on which we can obtain the optimal dual solution to (D1) as \(\lambda^*\), \(\nu^*\), and \(\boldsymbol{Z}^*\).

\subsection{Optimal Solution to Primal Problem (P1)}

Now, we present the optimal solution to primal problem (P1). With optimal dual variables \(\lambda^*\), \(\nu^*\), and \(\boldsymbol{Z}^*\) at hand, the corresponding unique optimal solution \(\boldsymbol{S}_{11}^*\) to problem \eqref{L_P_1} can be directly used for constructing the optimal primal solution to (P1), denoted by \(\boldsymbol{S}_{11}^{\mathrm{opt}}\). However, as indicated in Section IV-A, the optimal solutions of \(\boldsymbol{S}_{10}^*\) and \(\boldsymbol{S}_{00}^*\) to \eqref{L_P_1} are not unique. As a result, with given \(\boldsymbol{S}_{11}^{\mathrm{opt}}\), we need to find the optimal solutions of \(\boldsymbol{S}_{10}\) and \(\boldsymbol{S}_{00}\), denoted by \(\boldsymbol{S}_{10}^{\mathrm{opt}}\) and \(\boldsymbol{S}_{00}^{\mathrm{opt}}\), by solving one additional feasibility problem. We have the following proposition.

\begin{proposition} \label{prop2}
	The optimal solution to problem (P1) is
	\begin{equation} \label{P1_opt}
		\boldsymbol{S}^{\mathrm{opt}} = \begin{bmatrix}
			\boldsymbol{Q}^{\overline{\mathrm{null}}} \ \boldsymbol{Q}^{\mathrm{null}}
		\end{bmatrix} \begin{bmatrix}
			\boldsymbol{S}_{11}^{\mathrm{opt}} & \boldsymbol{S}_{10}^{\mathrm{opt}} \\ {\boldsymbol{S}_{10}^{\mathrm{opt}}}^H & \boldsymbol{S}_{00}^{\mathrm{opt}}
		\end{bmatrix} \begin{bmatrix}
		\boldsymbol{Q}^{\overline{\mathrm{null}}} \ \boldsymbol{Q}^{\mathrm{null}}
	\end{bmatrix}^H,
	\end{equation}
	where \(\boldsymbol{S}_{11}^{\mathrm{opt}}\) is given by \eqref{S_11} based on \(\lambda^*\), \(\nu^*\), and \(\boldsymbol{Z}^*\), and \(\boldsymbol{S}_{10}^{\mathrm{opt}}\) and \(\boldsymbol{S}_{00}^{\mathrm{opt}}\) are obtained by solving the following feasibility problem.
	\begin{equation}
		\begin{aligned}
			\mathrm{find} &\ \boldsymbol{S}_{10} \ \text{and} \ \boldsymbol{S}_{00} \\
			\mathrm{s.t.} &\ \eqref{Ra}, \ \eqref{Rc}, \ \eqref{Rb-reformulation}, \ \text{and} \ \eqref{S_p}.
		\end{aligned}
	\end{equation}
\end{proposition}

\begin{remark} \label{rem1}
	We have the following interesting observations from Proposition \ref{prop2}. First, the optimal transmit covariance solution \(\boldsymbol{S}^{\mathrm{opt}}\) follows the EMT structure based on the composite channel \(\boldsymbol{D}\) consisting of ID, EH, and sensing channels (see \eqref{D}), together with the water-filling-like power allocation (see \eqref{power_allo}). Next, it is observed that \(\boldsymbol{S}^{\mathrm{opt}}\) is divided into two parts, i.e., \(\boldsymbol{S}_{11}^{\mathrm{opt}}\) for the triple roles of communication, sensing, and powering; and \(\boldsymbol{S}_{10}^{\mathrm{opt}}\) and \(\boldsymbol{S}_{00}^{\mathrm{opt}}\) for sensing and powering only. Notice that based on extensive simulations, for the cases with randomly generated channels, we have \(\boldsymbol{D} \succ \boldsymbol{0}\) and \(\boldsymbol{S}^{\mathrm{opt}} = \boldsymbol{Q}^{\overline{\mathrm{null}}} \boldsymbol{S}_{11}^{\mathrm{opt}} {\boldsymbol{Q}^{\overline{\mathrm{null}}}}^H\), i.e., only \(\boldsymbol{S}_{11}^{\mathrm{opt}}\) is needed; by contrast, in some special cases (e.g., \(\boldsymbol{H}_{\text{ID}}\) and \(\boldsymbol{H}_{\text{EH}}\) being orthogonal to \(\boldsymbol{A}\)), it could happen that \(\boldsymbol{D}\) is rank-deficient, and \(\boldsymbol{S}_{10}^{\mathrm{opt}}\) and \(\boldsymbol{S}_{00}^{\mathrm{opt}}\) are also needed.
\end{remark}

\section{Numerical Results}

This section evaluates the performance of our proposed optimal design. In the simulation, the H-AP is equipped with a ULA of \(M = 10\) and \(N_{\text{S}} = 16\) antennas with half-wavelength spacing between consecutive antennas. The target angle is \(\theta = \frac{\pi}{3}\), and the reflection coefficient is set as \(\alpha = 10^{-8}\), accounting for a round-trip path loss of \(160\) dB. The ID channel and the EH channel are set as \(\boldsymbol{H}_{\text{ID}} = \alpha_{\text{ID}} \widehat{\boldsymbol{H}}_{\text{ID}}\) and \(\boldsymbol{H}_{\text{EH}} = \alpha_{\text{EH}} \widehat{\boldsymbol{H}}_{\text{EH}}\),
where \(\alpha_{\text{ID}}^2\) and \(\alpha_{\text{EH}}^2\) correspond to the path loss of \(120\) dB and \(60\) dB, respectively, and \(\widehat{\boldsymbol{H}}_{\text{ID}}\) and \(\widehat{\boldsymbol{H}}_{\text{EH}}\) correspond to the normalized channel accounting for the small-scale fading. The transmission frame length is set as \(L = 256\). The noise powers at the sensing receiver and the ID receiver, i.e., \(\sigma_{\text{S}}^2\) and \(\sigma_{\text{ID}}^2\), are set to be \(-80\) dBm.

\begin{figure}[tb]
	\centering 
	{\includegraphics[width=0.32\textwidth]{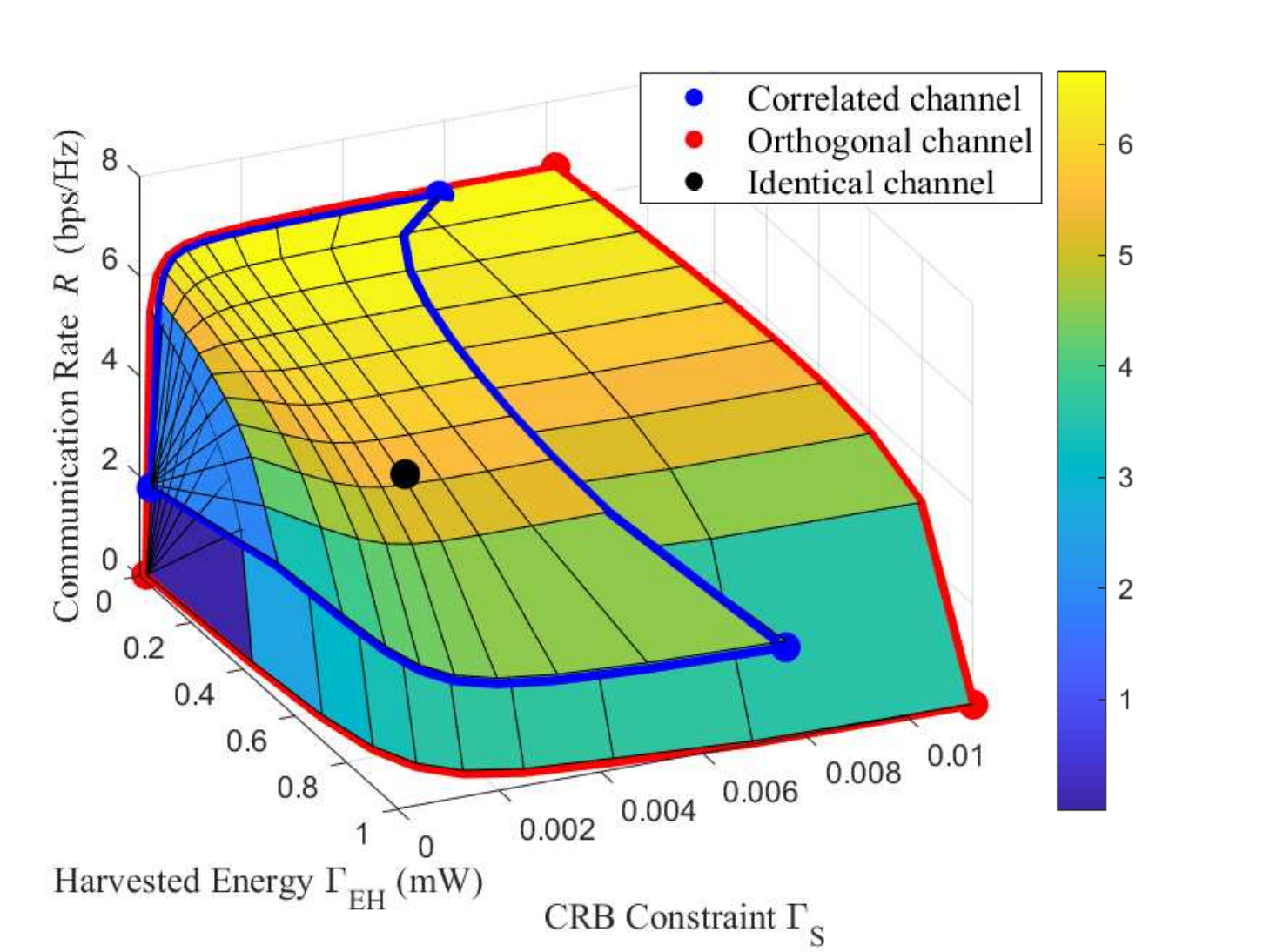}} 
	\caption{Pareto boundary for the C-R-E region with \(N_{\text{ID}} = N_{\text{EH}} = 1\) and \(P = 50\) dBm.}
\end{figure}
Fig. 3 shows the Pareto boundary of the C-R-E region with \(N_{\text{ID}} = N_{\text{EH}} = 1\) and \(P = 50\) dBm. In this figure, we consider the line-of-sight (LoS) channels for the ID and EH channels, i.e., \(\widehat{\boldsymbol{H}}_{\text{ID}} = \boldsymbol{a}_t^T(\theta_{\text{ID}})\) and \(\widehat{\boldsymbol{H}}_{\text{EH}} = \boldsymbol{a}_t^T(\theta_{\text{EH}})\). Furthermore, we set \(\theta = 0\),  \(\sin \theta_{\text{ID}} = \frac{2 \gamma}{M}\), and \(\sin \theta_{\text{EH}} = \frac{4 \gamma}{M}\). We consider three cases with $\gamma = 0$, $0.4$, and $1$, which correspond to the cases when the sensing, ID, and EH channels are identical, correlated, and orthogonal, respectively. It is observed that when $\gamma = 0$, the Pareto boundary is a point, which means that the R-max, E-max, and C-min strategies are identical, and thus the three performance metrics are optimized at the same time. In is also observed that when $\gamma=1$, optimizing one metric (e.g., E-max) leads to poor performances for the other two metrics (e.g., zero rate and highest CRB), thus showing that the three objectives are competing. Furthermore, when $\gamma =0.4$, the C-R-E region boundary is observed to lie between those with $\gamma = 1$ and $\gamma = 0$, thus showing the C-R-E tradeoff when the channels are correlated. 

Next, we evaluate the performance of our proposed design as compared to the benchmark scheme based on time switching. In this scheme, the transmission duration is divided into three portions, \(t_{\text{ID}} \ge 0\), \(t_{\text{EH}} \ge 0\), and \(t_{\text{S}} \ge 0\), with \(t_{\text{ID}} + t_{\text{EH}} + t_{\text{S}} = 1\), during which the H-AP employs the transmit covariance matrices \(\boldsymbol{S}_{\text{ID}}\), \(\boldsymbol{S}_{\text{EH}}\), and \(\boldsymbol{S}_{\text{S}}\), respectively. The corresponding communication rate, harvested energy, and estimation CRB in this scheme are expressed as \({R}_{\text{TS}}(t_{\text{ID}}, t_{\text{EH}}, t_{\text{S}}) = t_{\text{ID}} {R}(\boldsymbol{S}_{\text{ID}})\), \({E}_{\text{TS}}(t_{\text{ID}}, t_{\text{EH}}, t_{\text{S}}) = t_{\text{ID}} {E}(\boldsymbol{S}_{\text{ID}}) + t_{\text{EH}} {E}(\boldsymbol{S}_{\text{EH}}) + t_{\text{S}} {E}(\boldsymbol{S}_{\text{S}})\), and \(\mathrm{CRB}_{\text{TS}}(t_{\text{ID}}, t_{\text{EH}}, t_{\text{S}}) = \mathrm{CRB}(t_{\text{ID}} \boldsymbol{S}_{\text{ID}} + t_{\text{EH}} \boldsymbol{S}_{\text{EH}} + t_{\text{S}} \boldsymbol{S}_{\text{S}})\), respectively. Then, we optimize \(t_{\text{ID}}\), \(t_{\text{EH}}\), and \(t_{\text{S}}\) to achieve different C-R-E tradeoffs.

\begin{figure}[tb]
	\centering 
	{\includegraphics[width=0.32\textwidth]{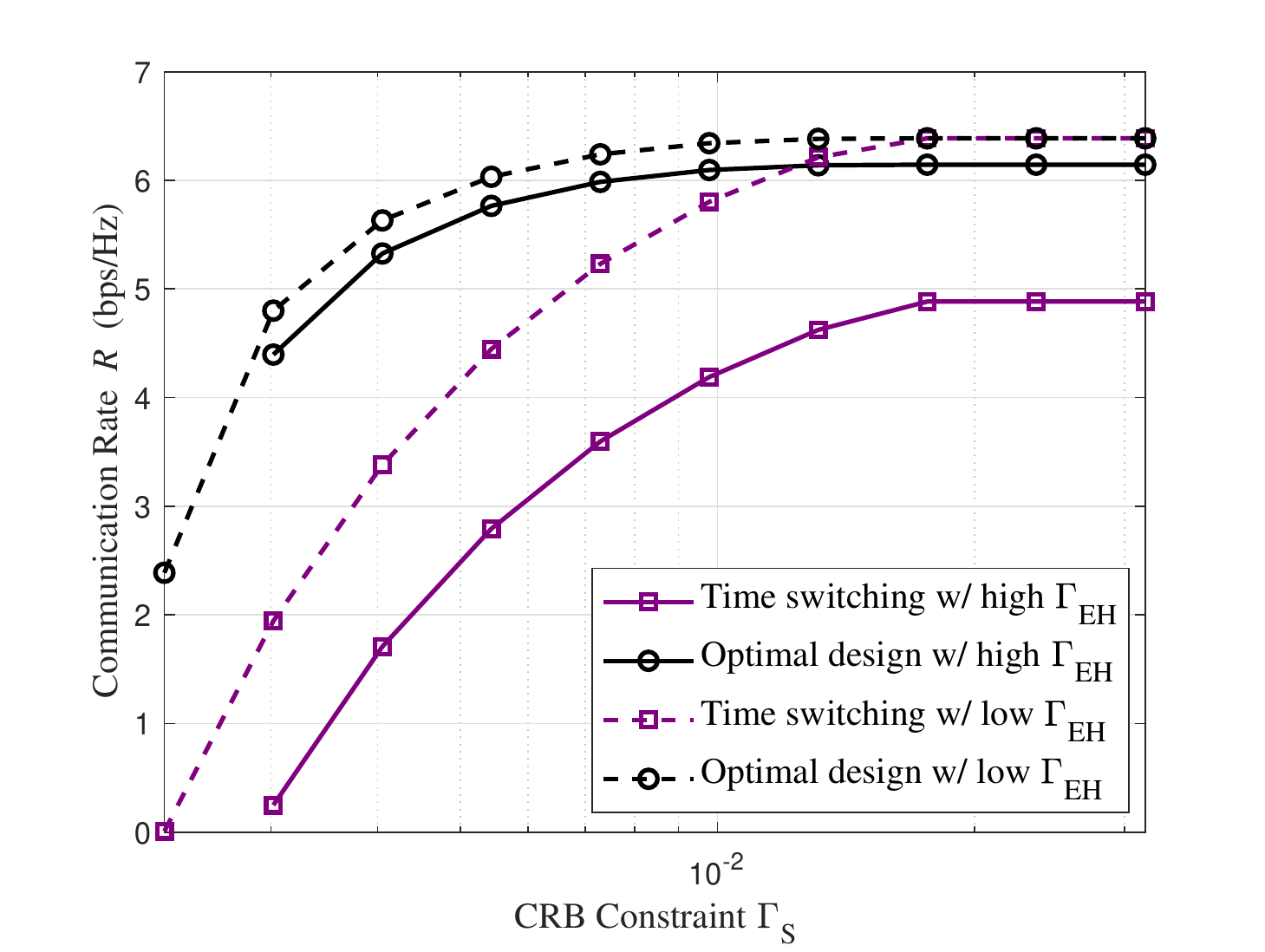}} 
	\caption{The communication rate \({R}\) versus the CRB constraint \(\Gamma_{\text{S}}\) with \(N_{\text{ID}} = N_{\text{EH}} = 4\) and \(P = 40\) dBm.} 
\end{figure}
Fig. 4 shows the obtained communication rate \({R}\) by (P1) versus the estimation CRB constraint \(\Gamma_{\text{S}}\), in which two EH constraints \(\Gamma_{\text{EH}} = 0.5 {E}_{\mathrm{max}}\) (high \(\Gamma_{\text{EH}}\) case) and \(\Gamma_{\text{EH}} = 0.05 {E}_{\mathrm{max}}\) (low \(\Gamma_{\text{EH}}\) case) are considered. Here, we set \(N_{\text{ID}} = N_{\text{EH}} = 4\) and \(P = 40\) dBm, and generate \(\widehat{\boldsymbol{H}}_{\text{ID}}\) and \(\widehat{\boldsymbol{H}}_{\text{EH}}\) as CSCG random matrices with each element being zero mean and unit variance. It is observed that under each EH constraint, our proposed optimal design outperforms the time switching scheme. For the case when \(\Gamma_{\text{EH}}\) is high, the performance gap is observed to be significant over the whole regime of \(\Gamma_{\text{S}}\). By contrast, for the case when \(\Gamma_{\text{EH}}\) is low, the time switching scheme is observed to perform close to the optimal design when \(\Gamma_{\text{S}}\) becomes large. This is due to the fact that both schemes can achieve the R-max vertex with the maximum communication rate in this case.

\section{Conclusion}

This paper studied the fundamental C-R-E tradeoff performance limits for a new multi-functional MIMO system integrating triple functions of sensing, communication, and powering. We characterized the Pareto boundary of the so-called C-R-E region, by optimally solving a new MIMO communication rate maximization problem subject to both energy harvesting and estimation CRB constraints. It was shown that the resultant C-R-E tradeoff highly depends on the correlations of the ID, EH, and sensing channels, and the proposed optimal design significantly outperforms the benchmark scheme based on time switching. 

\bibliographystyle{IEEEtran}
\bibliography{IEEEabrv,SWIPTISAC}

\end{document}